\documentclass[11pt,english]{article}
\usepackage{graphicx}
\usepackage{amstext}
\usepackage{caption}
\usepackage{etoolbox}
\usepackage{makeidx}
\include{epsf}
\usepackage{amsfonts}
\usepackage{authblk} 
\usepackage{sectsty}
\usepackage{amsmath,amssymb,epsfig}
\usepackage{amscd}
\usepackage{amsthm}
\usepackage{mathrsfs}
\usepackage{dsfont}
\usepackage[applemac]{inputenc}
\usepackage[english]{babel}
\usepackage{enumitem} 
\usepackage[]{latexsym}
\usepackage{braket}
\usepackage{caption}

\usepackage{titlesec}
\usepackage[titletoc,toc,title]{appendix}

\newtheorem{theorem}{Theorem}[section]	
\newtheorem{proposition}{Proposition}[section]
\titleformat{\section}{\normalfont\Large\bfseries}{\thesection.}{0.5em}{}
\titleformat{\subsection}{\normalfont\itshape}{\thesubsection.}{0.5em}{}
\titleformat{\subsubsection}{\normalfont\itshape}{\thesubsubsection.}{0.6em}{}

\usepackage{hyperref}
\hypersetup{
pdftitle={},%
pdfauthor={},%
pdfsubject={},%
pdfkeywords={},%
colorlinks=true,%
linkcolor=blue,%
citecolor=red,%
linktocpage=true,%
hyperfootnotes=true,%
pageanchor=true
}

\newcommand{\be}{\begin{equation}}
\newcommand{\ee}{\end{equation}}
\def\beqa{\begin{eqnarray}}
\def\eeqa{\end{eqnarray}}
\def\bean{\begin{eqnarray*}}
\def\eean{\end{eqnarray*}}

\newcommand{\eqn}[1]{(\ref{#1})}

\newcommand{\Tr}[1]{\:{\rm Tr}\,#1}

\renewenvironment{thebibliography}[1]
         {\section*{References}\frenchspacing\small
          \begin{list}{[\arabic{enumi}]}
         {\usecounter{enumi}\parsep=2pt\topsep 0pt
         \settowidth{\labelwidth}{[#1]}
         \leftmargin=\labelwidth\advance\leftmargin\labelsep
         \rightmargin=0pt\itemsep=1pt\sloppy}}{\end{list}}

 \numberwithin{equation}{section}

\input xy
\xyoption{all}

\title{\textbf{Tomographic Reconstruction of  Quantum Metrics}\vspace{0.5cm}}
\date{}

\author[1]{Marco Laudato}
\author[1,2]{Giuseppe Marmo}
\author[1]{Fabio M. Mele}
\author[1,2]{Franco Ventriglia}
\author[1,2]{Patrizia Vitale}

\affil[ ]{}
\affil[1]{\textit{\footnotesize Dipartimento di Fisica ``E. Pancini'' Universit\`a di Napoli Federico II, Complesso Universitario di Monte S. Angelo, via Cintia, 80126 Naples, Italy.}}
\affil[2]{\textit{\footnotesize INFN, Sezione di Napoli, Complesso Universitario di Monte S. Angelo, via Cintia, 80126 Naples, Italy.}}
\affil[ ]{}
\affil[ ]{\footnotesize e-mail: \texttt{marcolaudato@hotmail.com, marmo@na.infn.it, mele.fabio91@gmail.com, ventriglia@na.infn.it, patrizia.vitale@na.infn.it}}

\begin{document}

\maketitle

\begin{abstract}
\small
In the framework of quantum information geometry we investigate the relationship between monotone  metric tensors uniquely defined on the space of quantum tomograms, once the tomographic scheme chosen, and monotone quantum metrics on the space of quantum states, classified  by operator monotone functions, according to Petz classification theorem. We show  that different  metrics can be related through a change of  the tomographic map  and prove that there exists a bijective relation between monotone quantum metrics associated with  different operator monotone functions. Such bijective relation    is uniquely defined in terms of solutions of a first order second degree differential equation for the parameters of the involved tomographic maps.  We first exhibit an example of a non-linear tomographic map which connects a monotone metric with a new one which is not monotone. Then we provide a second example where  two monotone metrics are uniquely related through their tomographic parameters. 
\end{abstract}

\newpage

\section{Introduction}	
A considerable effort has been devoted in recent years  to understand the role of Information Theory in Physics. In particular, due to the common probabilistic background, the study of Quantum Mechanics from the point of view of Information Theory has been very fruitful and it allowed the foundation of a brand-new field called Quantum Information Theory \cite{Chuang}. It is now a well-established subfield of physics, having relevance also to other areas of science.

Starting from the pioneering work of  Cramer, Rao and Fisher in the 40's,   Amari and others (see \cite{Am85} and references therein) have shown that it is possible to approach Information Theory by using tools from Differential Geometry. Specifically, Geometric Information Theory deals with a Riemannian manifold of probability distributions on some sample space $X$ endowed with a pair of dually related connections. The Riemannian structure can be defined in terms of the Hessian of a class of functions called \textit{potential functions}, or alternatively \textit{contrast functions}, \textit{divergence functions} or \textit{distinguishability functions}. They are defined on two copies of the statistical manifold \footnote{A dynamical framework where two copies of the statistical manifold are connected with the tangent bundle by means of the Hamilton-Jacobi theory can be found in \cite{CDFMMP} .} and, roughly speaking, give a measure of the distance between two probabilities. From the Chentsov uniqueness theorem \cite{1} we know that, in the classical case at finite dimensions, there exists a unique monotone metric on  statistical manifolds  usually called Fisher-Rao metric tensor \cite{FR} (the generalization to the infinite-dimensional case was proven by Michor in \cite{michor}).

On the other hand, in the quantum setting, the probabilistic interpretation of Quantum Mechanics deals with probability amplitudes instead of probabilities. It has been shown \cite{GQM9} that, by replacing probabilities with probability amplitudes, the Fisher-Rao metric tensor used in classical Information Theory may be obtained from the Fubini-Study metric on the space of pure states of a quantum system. Moreover, in order to include also mixed states, it is possible to``quantize" the simplex of classical  probabilities by associating with every probability vector a coadjoint orbit of the unitary group acting on the dual space of its Lie algebra \cite{Erc1,Erc2}. The union of these orbits can be identified with the stratified manifold of the quantum state space \cite{GKM}. It is possible to consider this space as a \textit{quantized} statistical manifold and, by means of a potential function\footnote{To this aim, different quantum entropies have been suggested. The most celebrate is the von Neumann entropy, but also other candidates, like Tsallis or the R\'enyi's entropy, have been considered.} it is possible to define quantum Fisher metric tensors on it. However, unlike the classical case, these metric tensors are far from being  unique. They were classified by Petz \cite{2} in terms of operator monotone functions. In particular, as we will recall soon, the Petz classification theorem states that there exists a bijective correspondence between quantum metrics and operator monotone functions $f:(0,\infty)\rightarrow\mathbb R$ which satisfy  $f(t)=tf(1/t)$.

An alternative approach to conventional pictures of quantum mechanics is provided by the tomographic picture. Here a quantum state is associated with a fair probability distribution, a tomogram, and each probability distribution depends on the state and on a resolution of the identity. When a sufficient set of resolutions are used (a \textit{quorum}) it is possible to reconstruct the state from the various probability distributions associated with it (for a recent review see \cite{introtom,pedatom} and refs. therein).
Since tomographic probability distributions are fair probabilities, Chentsov's theorem ensures that there is a unique tomographic metric defined on them. It has been shown in \cite{3} that, starting from the tomographic metric,  it is possible to reconstruct  the quantum metric obtained from the Tsallis entropy on the  space of states.  
Here the main point emerges. Since from the usual spin-tomography \cite{DodPLA, OMJETP, MarPhSc00, Ventrrev}, we can obtain \textit{one} metric tensor  which is in turn associated by Petz classification theorem to \textit{one} operator monotone function, it is clear that a choice has been made  somewhere to select a unique metric in the family of quantum metrics.
Our claim is that this choice is actually the choice of the particular tomographic scheme we have used. Moreover, since the tomographic  metric is uniquely  determined by the monotonicity requirement, it is natural to enquire about the monotonicity of the quantum metric that is reconstructed.

The scope of the present work is to investigate the relationship between  the choice of tomographic schemes and  
monotone operator  functions classifying   
quantum metrics.    In order to avoid computational difficulties, we shall eventually  restrict the analysis to qubits. 

The paper  is organized as follows. In Section \ref{tomography} we review the derivation of the tomographic metric \cite{3} from a divergence function, in the spin-tomographic scheme. In Section \ref{cts} we analyze the meaning of a change of tomography from a geometrical point of view. We then discuss   the relationship between the metric tensors defined on the space of tomographic probability distributions, before and after the change of the tomographic scheme. Then in Section \ref{states} we discuss the corresponding transformation induced on the metric defined on the space of quantum states. Specifically, in Section \ref{secnl} we exhibit an example of a metric arising from a new tomographic scheme, non-linearly related to the usual spin-tomographic one, which is, however, not monotone. 
Secondly we  derive   an invertible relation between the tomographic parameters of related operator monotone metrics in terms of the solution of a  first order, second degree, ordinary differential equation.  Finally we find a solution providing an example of this proposition. Some of the calculations  are reported in Appendix \ref{A}.

\section{Spin-tomographic metric from divergence functions}\label{tomography}

In the spin-tomographic setting \cite{DodPLA, OMJETP, MarPhSc00, Ventrrev},  a $N$-level quantum state  represented by   a density matrix, $\rho$,   is   associated with  a  tomographic probability distribution (a \textit{tomogram}) $\mathcal{W}(m|u)$, through a choice of a dequantizer operator, $D$, according to the general dequantization  procedure \cite{MMV1, MMV2} $\rho\mapsto\mathcal{W}= \Tr \rho D$.  The tomography is realized by unitarily rotating the spin basis in the space of quantum states:  by posing $N=2j+1 $, with $j$ the quantum number associated to the spin  and $-j\le m\le j$,   the operator  $D(m,u)=u^\dag |m\rangle\langle m| u\ $,  is  the spin-tomographic family of dequantizers,  depending on $m$ and  $u \in SU(N)$. 
We have indeed a resolution of the identity for each $u$ 
\be
\mathds1_u=\sum_m u  |m\rangle\langle m|u^\dag=  \sum_m |m, u\rangle\langle m, u|
\ee
with $|m,u\rangle$  providing a basis in the unitarily rotated reference frame in the Hilbert space of states and  the unitary matrix $u$ labelling  the reference frame where the spin states are considered. 
The spin-tomographic probability distribution is then given by the diagonal matrix elements of $u\rho u^\dagger$, i.e. the density operator calculated in the given reference frame associated to $u$, that is
\begin{equation}
\label{tomografia}
\mathcal W_\rho(u|m)=\Tr\bigl(\rho u^\dagger\ket{m}\bra{m}u\bigr)=\bra{m}u\rho u^\dagger\ket m.
\end{equation}
Eq. \eqn{tomografia} is invertible if a sufficient number of reference frames,  called {\it quorum}, is provided. As discussed in \cite{3}, a sufficient number of frame is $N+1$. 

According to \cite{Am85} we  can define a Fisher metric tensor on the space of (classical) tomographic probabilities starting from divergence functions. As in \cite{3}   we use 
 the  {\it relative Tsallis entropy}
 \be
0\le S_q(\mathcal \rho, \tilde\rho, u)= \left[(q(1-q)\right]^{-1}\left(1-  \sum_m {\mathcal{W}_\rho}^q(m |u) \,  {\widetilde{\mathcal{W}}_{\tilde\rho}}^{1-q}(m | u)\right)
\ee
with  $\rho, \tilde\rho$ corresponding to two different states,  and  $u$ labelling the reference frame.  The metric is then \cite{3} 
\be
G=-i^*d\, \tilde d S_{q}(\mathcal{W}_\rho, {\widetilde{\mathcal{W}}_{\tilde\rho}})=\left(q(1-q)\right)^{-1}\sum_m d  {\mathcal{W}_\rho}^q \otimes d  {\mathcal{W}_{\rho}}^{1-q}=  \sum_m \mathcal{W}_\rho d\ln\mathcal{W}_\rho \otimes d\ln\mathcal{W}_\rho \label{tomomet}
\ee
with $i: M\hookrightarrow M\times M$ the diagonal embedding of the parameters manifold $M$ into the Cartesian product, $i^*$ its pull-back and $d$,  $\tilde d$, exterior derivatives on $M\times M$ respectively acting on the left and  right components  of the Cartesian product. Note that in (\ref{tomomet}) there is no dependence on the parameter $q$ in agreement with what has already been pointed out in the introduction that, being tomograms fair probability distributions, Chentsov's theorem ensures the existence of a unique monotone metric defined on them. As expected, this has actually the form of the Fisher-Rao metric on the space of tomographic probabilities.

Let us restrict to the qubit case, where 
\be
\rho= \frac{1}{2}(\sigma_0 + \sum_jy_j \sigma_j) \label{qubit}
\ee
$\sigma_0, \sigma_j, j=1,..,3$ being the generators of the $\mathfrak{u}(2)$ algebra and  $\sum_ky_k^2=w^2\le 1$.  In this case the quorum is equal to $N+1=3$ and we can choose for example
\be
u_{1}= \exp(i \frac{\pi}{4}\sigma_2)\quad,\quad u_{2}= \exp(-i \frac{\pi}{4}\sigma_1)\quad,\quad u_{3}= {\mathds1} \label{refre}
\ee
so to have
\be
\mathcal{W}(\pm \frac{1}{2}|u_{j})\equiv \langle \pm\frac{1}{2}|u_{j} \rho u_{j}^\dag|   \pm\frac{1}{2} \rangle = \frac{1\pm y_j}{2}\;\;\; j=1,...,3
\ee
with $\mathcal{W}(\frac{1}{2}|u_{j})+ \mathcal{W}(-\frac{1}{2}|u_{j})=1$ for each $j$. 
On inverting this relation for the parameters $y_j$ we thus get an expression of the qubit state \eqn{qubit} in terms of tomograms $\mathcal{W}_j\equiv\mathcal{W}(\frac{1}{2}|u_{j})$
\be
y_j= 2 \mathcal{W}_j-1 \label{invers}
\ee
For each reference frame of the tomographic procedure, labelled by $u_j$ (cfr. \eqn{refre}), we get from   \eqn{tomomet} a symmetric tomographic tensor, which  can be expressed in terms of the parameters $y_j$ 
\be
G_{\mathcal W}  (y, u_j)= G_{rs}(y,u_j) dy_r\otimes dy_s \label{tomometric}
\ee
with 
\be
G_{rs}  (y,u_j)= \frac{1}{4}\frac{1}{\mathcal{W}_j(1-\mathcal{W}_j)} \delta_{rs}\delta_{sj}=\frac{1}{1-y_j^2} \delta_{rs}\delta_{sj}
\label{Gjk}
\ee
from which we get
\be
\mathcal{W}_j=\frac{1\pm \sqrt{1-1/G_{jj}}}{2}
\ee
Notice that the tensor \eqn{tomometric} is degenerate. 
Finally, on using Eq. \eqn{invers} , we arrive at a direct expression of the parameters $y_k$ labelling the quantum state, in terms of  the tomographic tensor
\be
y_j= \pm\sqrt{1-G_{jj}^{-1}}. \label{yjG} 
\ee
It is now possible to assemble together a sufficient number of these symmetric tomographic tensors so as to build a metric tensor on the space of states. This will be the subject of next section.
\subsection{Quantum metrics and reconstruction formula}
A  family of metric tensors on the manifold  of quantum states, $M$, has been obtained  in \cite{3} starting from the associated quantum relative Tsallis entropy
\be
{S}_{Ts}(\rho,\tilde\rho)= (1-\Tr\rho^q\tilde\rho^{1-q})[q(1-q)]^{-1} \label{Tsal}
\ee
The density matrices $\rho, \tilde \rho$ can be parametrized  in terms of diagonal matrices and  unitary transformations
\be
\rho= U\rho_0 U^{-1}, \;\;\; \tilde \rho= V\tilde\rho_0 V^{-1} \label{dmatrix}
\ee
where $ U,V\in SU(N)$ are special unitary $N\times N$ matrices, with $N$ labeling the levels. 
According to \cite{3} we have
\be
g=-i^*d\, \tilde d S_{Ts}(\rho,\tilde\rho)=\left(q(1-q)\right)^{-1}  i^{*}\Tr d\rho^q\otimes \tilde d\tilde \rho^{1-q} \label{genmet}
\ee
where the tensor product is to be understood in the space of one-forms,  whereas the trace is defined on the space of  $N\times N$ matrices. 
On using
\beqa
d \rho^q&=& d(U\rho_0^q U^{-1})=dU\rho_0^q U^{-1}+ U d\rho_0^q U^{-1}- U\rho_0^q U^{-1}d U U^{-1} \\
\tilde d \tilde\rho^{1-q}&=& d(V\tilde\rho_0^{1-q} V^{-1})=dV\tilde\rho_0^{1-q} V^{-1}+ V d\tilde\rho_0^{1-q} V^{-1}- V\tilde\rho_0^{1-q} V^{-1}d V V^{-1}
\eeqa
and performing the pullback to the manifold M (which amounts to put $\tilde\rho_0=\rho_0$ and $U=V, U^{-1}=V^{-1}$), 
we get\footnote{Note that the $q$-dependence, which disappears at the classical level as pointed out after Eq. (\ref{tomomet}), reappears very transparently at the quantum level (cfr. (\ref{generalg})).} \cite{3}
\be
g_q = \left(q(1-q)\right)^{-1} \Tr \Bigl( [U^{-1} dU, \rho_0^q]\otimes   [U^{-1} dU, \rho_0^{1-q}]+ q(1-q) \rho_0^{-1} d\rho_0\otimes d\rho_0 \Bigr).\label{generalg}
\ee
For $N=2$ the relation between the parametrization in \eqn{qubit}  and that in \eqn{dmatrix} acquires the simple form 
\be
y_j= \frac{1}{2} w \Tr (U\sigma_3 U^{-1} \sigma_j)\label{fondamentale}
\ee
whereas the metric \eqn{generalg} becomes
\begin{equation}
\label{Tmetric}
g_q=\frac{1}{1-w^2}dw\otimes dw+\frac{2}{q(1-q)} (a_q-b_q)(a_{1-q}-b_{1-q})(\theta_1\otimes\theta_1+\theta_2\otimes\theta_2), 
\end{equation}
with
\begin{equation}
a_\alpha=\left(\frac{1+w}{2}\right)^\alpha,\quad b_\alpha=\left(\frac{1-w}{2}\right)^\alpha,\quad\alpha=q,1-q, ~~~-1\le w\le 1
\end{equation}
and $\{\theta^k\}_{k=0,\dots,3}$ a basis of left-invariant 1-forms on $U(2)$ such that the left-invariant Maurer-Cartan 1-form can be written as $U^{-1}dU=i\sigma_k\theta^k$, with $\sigma_0$ the $2\times2$ identity matrix and $\sigma_j, j=1,2,3$, the Pauli matrices. In polar coordinates $y_1=w \sin\theta\cos\phi , \, y_2=w \sin\theta\sin\phi , \, y_3= w \cos\theta,$  the metric \eqn{Tmetric} reads
\be
g_q (w,\theta,\phi)= \frac{1}{1-w^2} d w\otimes dw + \frac{1}{2 q(1-q)}(a_q-b_q)(a_{1-q}-b_{1-q})\left(d\theta \otimes d\theta +\sin^2\theta d\phi\otimes d\phi \right)\label{giqu}
\ee
Thus,  thanks to Eq. (\ref{yjG}),  it is possible  \cite{3}  to reconstruct the quantum metric \eqn{giqu}  in terms of the tomographic tensors \eqn{tomometric}.   To summarize, although individual tomographic tensors are degenerate, a sufficient number of them allows to reconstruct the quantum metric if a quorum has been assigned.

\noindent
\textbf{Remark: }The metric (\ref{Tmetric}) is in agreement with  Petz classification theorem. Indeed, let $M_n(\mathbb C)$ be the space of all complex $n\times n$ matrices, and let $\mathcal D_n$ be the set of density matrices, i.e., the set of all positive definite $n\times n$ Hermitian matrices of trace 1. The tangent space $T_\rho\mathcal D_n$ at $\rho\in\mathcal D_n$ can be naturally identified with the set $\{A\in M_n(\mathbb C): A=A^\dagger, \text{Tr}A=0\}$. We recall that a linear mapping $\phi: M_n(\mathbb C)\rightarrow M_m(\mathbb C)$ is defined to be stochastic if $\phi(\mathcal D_n)\subseteq\mathcal D_m$. A metric $g_\rho$ is said to be monotone\footnote{Essentially, the monotonicity requirement is a quantum version of the so-called invariance criterion of classical information geometry (see \cite{Am85,1}), where classical stochastic maps are replaced with quantum stochastic maps.} if, for every stochastic map $\phi$, $g_{\phi(\rho)}(\phi(A),\phi(A))\leq g_\rho(A,A), \forall A\in T_\rho\mathcal D_n,\forall \rho\in\mathcal D_n$. 
A function $f$ is said to be operator monotone if, for $A, B$ Hermitean operators such that $A\le B$, $f(A)\le f(B)$.
\begin{theorem}\cite{2}(Petz classification theorem).
There exists a bijective correspondence between symmetric monotone metrics $g_\rho$ and symmetric operator monotone functions $f$, which is given by
\be
\label{petzth}
g_\rho(A,B)=\Tr\left(A ~ c_f(L_\rho,R_\rho)(B)\right)\qquad A,B\in T_\rho\mathcal D_n
\ee
where $c_f(x,y):=\frac{1}{yf\left(\frac{x}{y}\right)}, x,y>0$ is the so-called Chentsov-Morozova function associated to $f$, and $L_\rho(A):=\rho A, R_\rho(A):=A\rho$.
\end{theorem}
\noindent
For a two-level system the explicit expression of the correspondence (\ref{petzth}) is given by
\begin{equation}
\label{petz}
g_f=\frac{1}{1-w^2}dw\otimes dw+\frac{w^2}{(1+w)f(\frac{1-w}{1+w})}(\theta_1\otimes\theta_1+\theta_2\otimes\theta_2)\;,
\end{equation}
with $f$ the operator monotone function satisfying the symmetry property $f(t)=tf(1/t)$, and we pose $t=\frac{1-w}{1+w}$. Therefore, the metric (\ref{Tmetric}) can be obtained from the general formula (\ref{petz}) if we use as operator monotone function\footnote{Other examples of operator monotone functions and analysis of the associated metric tensors can be found in \cite{gib1, gib2}.}
\begin{equation}
\label{vecchiaomf}
f(t)=\frac{\bigl(q(1-q)\bigr)(t-1)^2}{(t^q-1)(t^{1-q}-1)}\qquad,\qquad t=\frac{1-w}{1+w}\,.
\end{equation}
With a slight abuse of notation, we will talk about {\it quantum metrics of Petz form} for all metrics that can be recast into the form given by  Eq. \eqn{petz}, regardless of $f$ being operator monotone or not. The only assumption will be that $f$ is a real analytic, positive function.

The following diagram summarizes the general setting we are dealing with:
\begin{equation}
\xymatrix{
\boxed{\text{Relative Entropy}}\ar[r]&\boxed{\text{Monotone Metric}}\ar[l]\ar[r]\ar[d]&\boxed{\text{Operator Monotone Function}}\ar[l]\\
&\boxed{\text{Tomography}}\ar[u]&
}
\end{equation}
Here the main point emerges. Since from the usual (linear) tomography (\ref{tomografia}) we can reconstruct only  \textit{one} quantum metric,  by means of the reconstruction formulae \eqn{yjG}, and such quantum metric  is in turn associated by Petz classification theorem to \textit{one} operator monotone function, it is clear that a choice must have been made somewhere to select a unique metric in the family of quantum metrics.

Our claim is that this choice is actually the choice of the particular tomographic scheme we have used.  Indeed, in general, one can define another (possibly non-linear) tomographic map as
\begin{equation}
\label{Ftomo}
\mathcal W_F(u|m)=\bra m uF(\rho)u^\dagger\ket m,
\end{equation}
where the function $F$ has to be invertible and such that $F(\rho)$ is still a quantum state.
In the forthcoming sections  we will show that, given two monotone metrics (namely two operator monotone functions),  they may be  related by a change of tomographic scheme of the form \eqn{Ftomo}, whose parameters will be determined in terms of solutions of an ordinary differential equation.

 To start with, let us  write the Fisher metric defined on tomographic probability distributions in the same  coordinates  as in  Eq. (\ref{petz}) .
  This  amounts to  locally rotate the Pauli matrices by means of  {\it dreibeine}  so to have 
\begin{equation}
\begin{cases}
\sigma_w=\sin\theta\cos\phi\,\sigma_1+\sin\theta\sin\phi\,\sigma_2+\cos\theta\,\sigma_3\\
\sigma_\theta=\cos\theta\cos\phi\,\sigma_1+\cos\theta\sin\phi\,\sigma_2-\sin\theta\,\sigma_3\\
\sigma_\phi=-\sin\phi\,\sigma_1+\cos\phi\,\sigma_2
\end{cases}
\end{equation}
or in matrix form
\be\label{polarsigma}
\sigma_w=\begin{pmatrix}\cos\theta&\sin\theta e^{-i\phi}\\ \sin\theta e^{i\phi}&-\cos\theta\end{pmatrix},\quad \sigma_\theta=\begin{pmatrix}-\sin\theta&\cos\theta e^{-i\phi}\\ \cos\theta e^{i\phi}&\sin\theta\end{pmatrix},\quad\sigma_\phi=\begin{pmatrix}0&-i\,e^{-i\phi}\\ i\,e^{i\phi}&0\end{pmatrix}
\ee
and write the tomograms in term of the eigeinstates of one of them. 
In particular, in what follows the reference frame in the Hilbert space of states will be given by the eigenstates $\{\ket{m_w}\}$ of $\sigma_w$. In this basis, the matrices (\ref{polarsigma}) assume the following expression
\be\label{dpolarsigma}
\sigma_w=\begin{pmatrix}1&0\\ 0&-1\end{pmatrix}\quad,\quad\sigma_\theta=\begin{pmatrix}0&e^{i\theta} e^{-i\phi}\\ e^{-i\theta} e^{i\phi}&0\end{pmatrix}\quad,\quad\sigma_\phi=\begin{pmatrix}0&-i\,e^{i\theta}e^{-i\phi}\\ i\,e^{-i\theta}e^{i\phi}&0\end{pmatrix}\;.
\ee
With respect to the basis $(\sigma_0,\sigma_\theta,\sigma_\phi,\sigma_w)$ of $\mathfrak{u}(2)$, the Bloch decomposition (\ref{qubit}) of a generic density matrix $\rho$ reads
\be
\label{polar}
\rho=\frac{1}{2}(\sigma_0+w\sigma_w)=\begin{pmatrix}\frac{1+w}{2}&0\\0&\frac{1-w}{2}\end{pmatrix}\;,
\ee
where we have used the relation $\sigma_w=\frac{\vec y\cdot\vec\sigma}{|\vec y|}=\frac{1}{w}\sum_{k=1}^3y_k\sigma_k$. In this case, a quorum is given by
\be
\label{wquorum}
u_\theta=\exp\left(i\frac{\pi}{4}\sigma_\phi\right)\quad,\quad u_\phi=\exp\left(-i\frac{\pi}{4}\sigma_\theta\right)\quad,\quad u_w=\mathds1
\ee
and the corresponding tomograms are
\be
\begin{split}
&\mathcal W_w\equiv\mathcal W_\rho(m_w=\pm1|u_w)=\braket{m_w|u_w\rho u_w^\dagger|m_w}=\frac{1\pm w}{2}\;,\\
&\mathcal W_\theta\equiv\mathcal W_\rho(m_w=\pm1|u_\theta)=\braket{m_w|u_\theta\rho u_\theta^\dagger|m_w}=\frac{1}{2}\;,\\
&\mathcal W_\phi\equiv\mathcal W_\rho(m_w=\pm1|u_\phi)=\braket{m_w|u_\phi\rho u_\phi^\dagger|m_w}=\frac{1}{2}\;.
\end{split}
\ee
Therefore, according to Eqs. (\ref{tomometric}, \ref{Gjk}), the unique non-zero tomographic tensor is the one associated to $u_w$, say
\be
\label{Gw}
G_w=G_{ww}\,dw\otimes dw=\frac{1}{1-w^2}\,d w\otimes dw\;.
\ee
Finally, by means of the relation
\be
w=\pm\sqrt{1-G_{ww}^{-1}}
\ee
it is possible to reconstruct the quantum metric (\ref{petz}) by means of the tomographic tensors.

\section{Changing the tomographic scheme}
\label{cts}
Let us analyze in details what the introduction of a new tomographic map  means and what are its main consequences.

A generic density state $\rho$ of a $N$-level quantum system can be parametrized in terms of diagonal matrices with positive entries and unitary transformations as
\begin{equation}\label{diag}
\rho(U,\vec p)=U\begin{pmatrix}
p_1 &\dots & 0\\
0 & p_2 & \dots\\
\dots & \dots &\dots\\
0 & \dots & p_N
\end{pmatrix} U^\dagger=U\rho_0U^\dagger\;,
\end{equation} 
where $U\in U(N)$ and, being $\text{Tr}\,\rho=1$, $\vec p\equiv(p_1,\dots,p_N)$ is a probability vector belonging to a $(N-1)$-simplex of classical probabilities. This essentially amounts to ``quantize'' the classical simplex by associating a coadjoint orbit of the unitary group with each probability vector identified with the diagonal elements of a density matrix \cite{3}. The union of all these orbits is the space of all quantum states. Clearly, the unitary matrices $U$ in (\ref{diag}) are determined only up to unitary transformations in the isotropy group $\mathbb G_0$ of $\rho_0$ and the parameter space $M$ is thus given by the homogeneous space $SU(N)/\mathbb G_0$ times the open part of the simplex to which the diagonal part of the state belongs. As stressed in \cite{3}, since such a homogeneous space is not parallelizable, we shall consider the differential calculus as carried on the redundant $SU(N)$ times the interior of the simplex.

At the tomographic level, the spin-tomographic probability distribution defined in (\ref{tomografia}) can be similarly parametrized by means of unitary matrices and diagonal matrices. Indeed, Eq. (\ref{tomografia}) can be recast in the form
\begin{equation}\label{tom}
\begin{split}
\mathcal W_{\rho}(u|m)&=\text{Tr}\bigl(u\rho u^\dagger\ket{m}\bra{m}\bigr)
=\text{Tr}\bigl((uU)\rho_0(uU)^\dagger\ket{m}\bra{m}\bigr)\\
&=\text{Tr}\bigl(v\rho_0v^\dagger\ket{m}\bra{m}\bigr)
\equiv\mathcal W_{\rho_0}(v|m)\;,
\end{split}
\end{equation}
where $U$ is the unitary matrix which diagonalizes $\rho$ and we have defined the unitary matrix $v$ as the product $uU$ with $u,U\in U(N)$. Hence, spin-tomograms come to be probability distributions defined on a discrete space $\mathcal X$ consisting of $2j+1$ points identified by the possible values of $m$ (e.g., two points for the qubit case) and, coherently with the parametrization (\ref{diag}) of the states, they are also parametrized in terms of $v\in SU(N)$ and $\rho_0$, namely $\mathcal W_{\rho_0}(v|m)\in\mathscr P(\mathcal X;M)$.

Let us consider now a change of the tomographic scheme, say
\begin{equation}\label{newtom}
\mathcal W_F(u|m)=\text{Tr}\bigl(uF(\rho)u^\dagger\ket{m}\bra{m}\bigr)\;,
\end{equation}
where $F$ is a real analytic, invertible function transforming states into states. The invertibility requirement ensures no loss of information about the starting state. Moreover, since $F$ is real analytic, $F(\rho)=F(U\rho_0U^\dagger)=UF(\rho_0)U^\dagger$, $U$ being the unitary matrix which diagonalizes $\rho$. Thus, Eq. (\ref{newtom}) can be written as
\begin{equation}
\mathscr P(\mathcal X;M)\ni\mathcal W_{F(\rho_0)}(v|m)=\text{Tr}\bigl(vF(\rho_0)v^\dagger\ket m\bra m\bigr)\;,
\end{equation} 
i.e., the spin-tomographic probability distributions associated with the new tomograms are still parametrized by elements in $M$. More specifically, the analyticity requirement implies that $F(\rho)$ is diagonalized by means of the same unitary matrix which diagonalizes  $\rho$ and therefore $F$ affects only the probability vector associated with the elements of $\rho_0$. Indeed, $F(\rho_0)$ is still a quantum state in its diagonal form and can be thus identified with a diagonal density matrix $\tilde\rho_0$ parametrized by a probability vector $\vec{\tilde p}$ depending on $\vec p$ through the following relation\footnote{Let us remark that Eq. (\ref{tilde}) implies equivariance with respect to unitary transformations.}
\begin{equation}\label{tilde}
F\bigl(\rho_0(\vec p)\bigr)=\tilde\rho_0(\vec{\tilde p})\;.
\end{equation}
In other words, the change of the tomographic scheme (\ref{newtom}) identifies a diffeomorphism on the parameter space $M$ acting non-trivially only on the simplex part, i.e.:
\begin{equation}\label{diffeo}
F\in {\rm Diff}(M)\qquad \text{s.t.}\qquad M\ni\bigl(\vec p,U\bigr)\longmapsto\bigl(\vec{\tilde p}(\vec p),U\bigr)\in M\\
\end{equation}

In what follows we will focus on two-level systems for which the unitary transformations diagonalizing the state $\rho$ belong to the group $SU(2)$ and the corresponding probability vector $\vec p$ can be parametrized as $\vec p=(p_1,p_2)=\bigl((1+w)/2,(1-w)/2\bigr)$ with $-1\leq w\leq 1$. In other words, in this case, the parameter space $M$ can be described in terms of $SU(2)$ times the open interval $(-1,1)$. 

Let us compute the spin-tomographic probability distribution associated with the state $\rho$ in the reference frame of eigenstates of $\sigma_w$, 
\be
\mathcal W_\rho(u|m_w)=\text{Tr}\left(u\rho u^\dagger\ket{m_w}\bra{m_w}\right)
\ee
with $\rho$ is given by (\ref{polar}). In such a reference frame, the change of the tomographic scheme (\ref{newtom}) reads
\be
\mathcal W_F(u|m_w)=\text{Tr}\bigl(uF(\rho)u^\dagger\ket{m_w}\bra{m_w}\bigr)=\text{Tr}\bigl(u\tilde\rho u^\dagger\ket{m_w}\bra{m_w}\bigr)=\mathcal W_{\tilde\rho}(u|m_w)\;,
\ee
with
\be
\tilde\rho=\frac{1}{2}(\sigma_0+\tilde w\sigma_w)\;.
\ee
Hence, Eq. (\ref{tilde}) relating the parameters of the states $\rho$ and $\tilde\rho$ now becomes
\be
F(\rho(w))=\tilde\rho(\tilde w)\;,
\ee
and we see that the change of tomography amounts to the following diffeomorphism on $M$ non-trivially acting only on the simplex part
\be
\label{wdiffeo}
M\ni\bigl(w,U\bigr)\overset{F}{\longmapsto}\bigl(\tilde w(w),U\bigr)\in M
\ee
We are interested in understanding how  tomographic tensors behave under a diffeomorphism of the form (\ref{wdiffeo}), i.e., under a change of the tomographic scheme. As already stressed in the introduction, quantum tomograms are true probability distributions, thus,  according to Chentsov's theorem, a unique monotone Fisher-Rao metric can be defined on their parameter space $M$ without ambiguities in complete analogy to the classical setting. 

According to Eq. \eqn{Gw},  the tomographic tensor acquires the following expression in terms of the parameter $\tilde w$ of $\tilde\rho$:
\begin{equation}\label{frtilde}
G_{\widetilde{\mathcal W}}=\sum_{m_w}\frac{1}{\mathcal W_{\tilde\rho}}\,d\mathcal W_{\tilde\rho}\otimes d\mathcal W_{\tilde\rho}=\frac{1}{1-\tilde{w}^2}\,d\tilde w\otimes d\tilde w\;.
\end{equation}
The pull-back along the map (\ref{wdiffeo}) thus yields
\begin{equation}
G_{\mathcal W_F}=F^*G_{\widetilde{\mathcal W}}=\frac{1}{1-\tilde{w}^2(w)}\,\left(\frac{d\tilde{w}(w)}{dw}\right)^2dw\otimes dw
=\frac{1-w^2}{1-\tilde w^2(w)}\,\left(\frac{d\tilde{w}(w)}{dw}\right)^2G_\mathcal W
\end{equation}
where $G_\mathcal W$  is given by (\ref{Gw}). Therefore, the action of the diffeomorphism $F$ on the tomographic metric essentially amounts to the following conformal transformation 
\begin{equation}\label{conformal}
G_{\mathcal W_F}=\mathcal A_FG_\mathcal W\;,
\end{equation}
where the conformal factor $\mathcal A_F$ is defined as
\begin{equation}\label{confo}
\mathcal A_F\equiv\mathcal A(\tilde w(w))=\frac{1-w^2}{1-\tilde w^2(w)}\left(\frac{d\tilde w(w)}{dw}\right)^2.
\end{equation}

\textbf{Remark:} Let us notice that the appearance of the conformal factor does not contradict Chentsov's theorem. Indeed, the latter states that the Fisher-Rao metric is the unique monotone metric invariant (up to a constant factor) under the action of stochastic maps, i.e., under diffeomorphisms on the space $\mathcal X$ of random variables \cite{1}, while in the case under examination the diffeomorphism $F$ is on the parameter space $M$ and not on $\mathcal X$.

To sum up, changing the tomographic scheme corresponds in general to perform a diffeomorphism of the form (\ref{diffeo}) on the parameter space whose action on the metric tensor defined on the space of tomographic probabilities maps Fisher-Rao metrics into conformal Fisher-Rao metrics, where the conformal factor is uniquely determined by the change of tomographic scheme.

\section{Counterpart on the space of states}
\label{states}
 
We are interested in characterizing the  correspondence between the choice of a tomographic scheme and operator monotone functions. To this aim we will study the form that the tomographic diffeomorphisms acquire on the space of quantum metrics.  In analogy with the classical case, we look for a  class of diffeomorphisms which map a quantum metric characterized by an operator monotone function to a different quantum metric characterized by another operator monotone function times a conformal factor which depends only on the change of tomographic scheme.

To explore the problem, we will first focus on the specific example of a non-linear exponential tomographic scheme, and then we will set up the general problem.

\subsection{An example of non-linear tomography for qubits} 		
\label{secnl}
In this section we will show by means of an explicit example that the diffeomorphism induced by changing the tomographic scheme maps the metric associated with  the von Neumann entropy to a different quantum metric which, up to the conformal factor  (\ref{confo}), is parametrized by a new function which is positive and real analytic on the interval $(0,\infty)$. After a deeper analysis we will find however that the function is  not operator monotone.
In particular, we will consider the tomography
\begin{equation}\label{newtom2}
\mathcal W_F(u|m_w)=\text{Tr}\bigl(uF(\rho)u^\dagger\ket{m_w}\bra{m_w}\bigr)\;,
\end{equation}
where we set
\begin{equation}
\label{exf}
F(\rho)=\frac{e^{-\beta\rho}}{Tr(e^{-\beta\rho})}.
\end{equation} 
Let us recall that, both being  states, we have the following Bloch decompositions for $\rho$ and $F(\rho)$
\begin{equation}
\label{fano}
\rho=\frac{1}{2}(\sigma_0+w\sigma_w)\quad,\quad F\bigl(\rho(w)\bigr)\equiv\tilde\rho(\tilde w)=\frac{1}{2}(\sigma_0+\tilde w\sigma_w)\qquad w, \tilde w\in(-1,1)
\end{equation}
which are diagonal in the basis of the eigenstates of $\sigma_w$. A straigthforward computation shows that the tomographic probabilities associated to the quorum (\ref{wquorum}), are given by
\begin{equation}
\begin{split}
&\mathcal W_{\rho}(\pm,u_\theta)=\mathcal W_{\rho}(\pm,u_\phi)=\frac{1}{2}\quad,\quad\mathcal W_{\rho}(\pm,u_w)=\frac{1\pm w}{2}\\
&\widetilde{\mathcal W}_{\tilde{\rho}}(\pm,u_\theta)=\widetilde{\mathcal W}_{\tilde{\rho}}(\pm,u_\phi)=\frac{1}{2}\quad,\quad\widetilde{\mathcal W}_{\tilde{\rho}}(\pm,u_w)=\frac{1\pm\tilde w}{2}.
\end{split}
\end{equation}
Therefore, with respect to the tomographic scheme (\ref{newtom2}), the only non-vanishing parameter of the state $\tilde\rho$ is given by the usual formula:
\begin{equation}
\tilde w=2\widetilde{\mathcal W}_{\tilde{\rho}}(+|u_w)-1.
\end{equation}
Comparing Eqs. (\ref{fano}), we obtain the relation $\tilde w=\tilde w(w)$ between the parameter of $\rho$ and the parameter of $\tilde\rho$. In the particular case of Eq. (\ref{exf}) we obtain\footnote{For the explicit computation we refer to Appendix \ref{A}.} the following relation:
\begin{equation}
\label{erho}
\tilde w(w)=-\tanh\left(\frac{\beta w}{2}\right).
\end{equation}
Since $\tilde\rho$ is still a quantum state, we can use, for example, as generating function of the quantum metric either the Von Neumann entropy
\begin{equation}
S_{vN}\bigl(\tilde\rho,\tilde\zeta\bigr)=\Tr \bigl(\tilde\rho(\ln \tilde\rho-\ln\tilde\zeta)\bigr)
\end{equation}
or the (rescaled) Tsallis entropy
\begin{equation}
S_T\bigl(\tilde\rho,\tilde\zeta\bigr)=\frac{1}{q(1-q)}\bigl(1-\Tr(\tilde\rho^q \;\tilde\zeta^{1-q})\bigr).
\end{equation}
{\bf Remark}. The von Neumann entropy can be obtained from the Tsallis entropy in the limit $q\rightarrow 1$. From them, we can compute the metric tensors
\begin{equation}
\label{VNmetric}
g(\tilde w)=\frac{1}{1-\tilde w^2}d\tilde w\otimes d\tilde w+2\tilde w\ln\left(\frac{1+\tilde w}{1-\tilde w}\right)(\theta_1\otimes\theta_1+\theta_2\otimes\theta_2)
\end{equation}
and
\begin{equation}
\label{Tmetric2}
g_q(\tilde w)=\frac{1}{1-\tilde w^2}d\tilde w\otimes d\tilde w+\frac{2}{q(1-q)}(a_q-b_q)(a_{1-q}-b_{1-q})(\theta_1\otimes\theta_1+\theta_2\otimes\theta_2)
\end{equation}
respectively. In \cite{3} it was proved that these two metrics are respectively related to the following operator monotone functions
\begin{equation}
f(t)=\frac{(t-1)}{\ln t}\quad,\quad f(t)=\frac{(q(1-q))(t-1)^2}{(t^q-1)(t^{1-q}-1)}
\end{equation}
where 
\begin{equation}
t=\frac{1-\tilde w}{1+\tilde w}.
\end{equation}

To prove our claim, we will proceed through the following two steps:
\begin{itemize}
\item[1)] We perform the pullback of the metric (\ref{VNmetric}) (or of the metric (\ref{Tmetric2})) with respect to the diffeomorphism (\ref{wdiffeo}) where the non-trivial part is given by Eq. (\ref{erho}). In this way we obtain a new tensor written in terms of the parameter $w$ of $\rho$;
\item[2)] We compare the coefficients of this new tensor with the general expression of quantum metric (\ref{petz}) written in terms of $w$, while letting unknown  the operator monotone function. If this  function exists, we have  shown that   the tomography (\ref{newtom2}) gives rise to a metric which is formally of the Petz type, however we are not guaranteed that the new metric is monotone. 
\end{itemize}
The first step\footnote{For simplicity, we will focus only on the metric obtained from the von Neumann entropy but the same considerations can be done also for the metric obtained from any other divergence function, e.g. the Tsallis entropy.} consists in replacing the expression (\ref{erho}) into the metric (\ref{VNmetric}):
\begin{equation}
\begin{split}
g\bigl(\tilde w(w)\bigr)&=\frac{1}{1-\tanh^2\left(\frac{\beta w}{2}\right)}\left(\frac{-\beta}{2\,\cosh^2\left(\frac{\beta w}{2}\right)}\right)^2dw\otimes dw+\\
&-2\,\tanh\left(\frac{\beta w}{2}\right)\ln\left(\frac{1-\tanh\left(\frac{\beta w}{2}\right)}{1+\tanh\left(\frac{\beta w}{2}\right)}\right)(\theta_1\otimes\theta_1+\theta_2\otimes\theta_2).
\end{split}
\end{equation}
In order to compare with the general expression (\ref{petz}), we need to factor out  a common factor $\bar{\mathcal A}(w)$ defined as:
\begin{equation}
\bar{\mathcal A}(w)=\frac{1-w^2}{1-\tanh^2\left(\frac{\beta w}{2}\right)}\left(\frac{-\beta}{2\,\cosh^2\left(\frac{\beta w}{2}\right)}\right)^2=\frac{\beta^2(1-w^2)}{4\,\cosh^2\left(\frac{\beta w}{2}\right)}.
\end{equation}
In this way the tensor factorizes as:
\begin{equation}
g_f\bigl(\tilde w(w)\bigr)=\bar{\mathcal A}(w)g_h\bigl(\tilde w(w)\bigr),
\end{equation}
where 
\begin{equation}
\begin{split}
&g_h\bigl(\tilde w(w)\bigr)=\frac{1}{1-w^2}dw\otimes dw+\\
&-2\,\tanh\left(\frac{\beta w}{2}\right)\ln\left(\frac{1-\tanh\left(\frac{\beta w}{2}\right)}{1+\tanh\left(\frac{\beta w}{2}\right)}\right)\frac{1-\tanh^2\left(\frac{\beta w}{2}\right)}{1-w^2}\left(\frac{4\,\cosh^4\left(\frac{\beta w}{2}\right)}{\beta^2}\right)(\theta_1\otimes\theta_1+\theta_2\otimes\theta_2).
\end{split}
\end{equation}
A straightforward computation shows that the conformal factor is exactly of the form (\ref{confo}), i.e., the same of the tomographic case, when the transformation $\tilde w(w)$ is given by Eq. (\ref{erho}). Therefore, if the analogy with the classical case holds, we have to check that the tensor $g_h\bigl(\tilde w(w)\bigr)$ can be recast in the form of the general metric (\ref{petz}). On using
\begin{equation}
\ln\left(\frac{1+x}{1-x}\right)=2\,{\rm Artanh}(x)
\end{equation}
we obtain the following expression for the candidate operator monotone function:
\begin{equation}
\label{tomf}
h\left(\frac{1-w}{1+w}\right)=\frac{\beta}{4}\frac{w(1-w)}{\sinh(\beta w)}.
\end{equation}
The symmetry property $t\,f(1/t)=f(t)$ can be checked by setting: 

\begin{equation}
t=\frac{1-w}{1+w}\quad\Longrightarrow\quad w=w(t)=\frac{1-t}{1+t}\,.
\end{equation}

\noindent
Indeed, the function (\ref{tomf}) becomes	

\begin{equation}
\label{ft}
h(t)=h\bigl(w(t)\bigr)=\frac{\beta}{4}\frac{1-t}{1+t}\frac{\left(1-\frac{1-t}{1+t}\right)}{\sinh\left(\beta\frac{1-t}{1+t}\right)}=\frac{\beta}{2}\frac{t(1-t)}{(1+t)^2\,\sinh\left(\beta\frac{1-t}{1+t}\right)}\,,
\end{equation}
and then
\begin{equation}
\label{f1t}
h\left(\frac{1}{t}\right)=-\frac{\beta}{2}\frac{1-t}{(1+t)^2}\frac{1}{\sinh\left(-\beta\frac{1-t}{1+t}\right)}=\frac{\beta}{2}\frac{1-t}{(1+t)^2\,\sinh\left(\beta\frac{1-t}{1+t}\right)}\,.
\end{equation}
Finally, by comparing equations (\ref{ft}) and (\ref{f1t}), we have that:
\begin{equation}
h(t)=th\left(\frac{1}{t}\right).
\end{equation}
In order to study the operator monotonicity of the function \eqn{ft} we resort to the following fundamental  theorem due to L\"owner \cite{lowner1,lowner2,lowner3}:
\begin{theorem}
Let $f\,:\,I\longrightarrow\mathbb R$ be a non-constant function  on an interval $I$ which is either finite or infinite. Then $f$ is operator monotone on $I$ iff it is real analytic on $I$  and  there exists  a holomorphic extension  to the open upper half-plane $\Pi_+$   such that $f(\Pi_+)\subseteqq\Pi_+$.
\end{theorem}
\noindent
As one can check, the imaginary part of the analytic extension of our candidate $h$ in Eq. \eqn{ft} to the upper complex plane is negative in a small neighbourhood of the point $z=-1$. Therefore the function \eqn{ft} is not operator monotone. This implies that the new tomographic scheme associated with $F$  generates a metric which is not monotone. 

\subsection{Tomography for Monotone Metrics}		
\label{secinverse}
We have seen that any change of tomography gives rise to a relation between the parameter of $\rho$, say $w$, and the parameter of $F(\rho)=\tilde\rho$, say $\tilde w$, that is $\tilde w=\tilde w(w)$. 
Upon replacing $\tilde w= \tilde w(w) $ the  quantum metric \eqn{petz} reads:
\begin{equation} \label{gf}
 g_{ f}\bigl(\tilde w(w)\bigr)=\frac{1}{1-\tilde w(w)^2}\left(\frac{d\tilde w(w)}{d w}\right)^2dw\otimes dw+ 
\frac{\tilde w(w) ^2}{(1+\tilde w(w))  f(\frac{1-\tilde w}{1+\tilde w})}(\theta_1\otimes\theta_1+\theta_2\otimes\theta_2).
\end{equation}
We then  factorize the pulled-back tensor as
\begin{equation} \label{Agh}
g_f\bigl(\tilde w(w)\bigr)=\mathcal A\bigl(\tilde w(w)\bigr)g_h\bigl(w\bigr),
\end{equation}
where
\begin{equation}
\mathcal A\bigl(\tilde w(w)\bigr)=\frac{(1-w^2)}{1-\tilde w(w)^2}\left(\frac{d \tilde w(w)}{d  w}\right)^2
\end{equation}
and
\begin{equation}
\label{Hmetric}
g_h(w)=\frac{1}{1-w^2}dw\otimes dw+\frac{w ^2}{(1+w)  h(\frac{1- w}{1+w})}(\theta_1\otimes\theta_1+\theta_2\otimes\theta_2).
\end{equation}
Notice that  $\mathcal A$ is exactly the conformal factor (\ref{confo}).  On comparing Eq. \eqn{gf} with  Eqs.   \eqn{Agh}, (\ref{Hmetric}), we obtain an equation for the unknown function $\tilde w (w)$
\begin{equation}\label{maineq}
\left(\frac{d \tilde w}{d w}\right)^2= \left(\frac{ \tilde w}{ w}\right)^2\frac{1-\tilde w}{1-w} ~ \frac{h(\frac{1- w}{1+w})}{f(\frac{1-\tilde w}{1+\tilde w})}
\end{equation}
which is a quadratic, first order differential equation.  Therefore, under the hypothesis of Lipschitzian continuity, a solution exists and it is unique. Thus we have proven the following  
\begin{proposition}\label{prop}
There is always a change of tomographic scheme which connects two given  metrics of the Petz form, characterized by real analytic, positive  functions $f$ and $h$. These metrics are monotone if and only if $f$ and $h$ are operator monotone functions on the same interval.  
\end{proposition}
To illustrate the result, let us consider the following example. Let $f(t)=  t^{2a}$, $h(t)= t^{2b}$, which are operator monotone functions on the interval $(0,\infty)$ for $0\le a,b\le 1/2$ \cite{bathia}. On separating variables in  Eq. \eqn{maineq} we have
\be
\frac{d \tilde w}{\tilde w(1-\tilde w)^{1/2-a}(1+\tilde w)^a}= \frac{d  w}{  w(1- w)^{1/2-b}(1+ w)^b}
\ee
The simplest choice is $a=1/2, b=0$ in which case we obtain
\be
{\rm Artanh}\sqrt{1-\tilde w}={\rm Artanh}\sqrt{1+w} . 
\ee
which gives $\tilde w(w)= -w$.     \\
{\bf Remark.} Let us notice that Eq. \eqn{maineq}  relates metrics of Petz form \eqn{petz} regardless of the operator monotonicity of $f$ and $h$. Indeed, the example considered in section \ref{secnl}  satisfies Eq. \eqn{maineq} with $f=\frac{(t-1)}{\ln t} $,  $h$ given in Eq. \eqn{f1t}, where $f$ is operator monotone but $h$ is not.

\section{Conclusions} 

Tomography deals with the reconstruction of quantum states when a sufficient (quorum) family of tomograms is available. A natural question arises in quantum information theory: is it possible to reconstruct quantum metrics from the knowledge of metrics on the space of tomograms (fair probability distributions)? An affirmative answer was provided in \cite{3}. A subsequent question arises also very natural. If we start with monotone metrics on the space of tomograms, are we going to obtain monotone quantum metrics out of the tomographic reconstruction?

In this paper we have considered this question by means of the classification provided by Petz in terms of operator monotone functions. By analyzing specific examples we have found that in some cases it is possible to obtain monotone metrics, provided our change of tomographic maps satisfies some nonlinear differential equation, however this condition alone is not sufficient to warrant the conclusion in all cases. In summary we have to conclude that a necessary and sufficient condition to
obtain monotone metric out of a nonlinear change of tomographic maps is still lacking. We hope to come back to this issue very soon.

The latter result has been derived for the qubit case. Let us notice that, for a generic invertible qudit state, the spectrum will be parametrized by $d-1$ parameters. Then, since the diffeomorphism  induced by a change of tomographic scheme is always of the form \eqn{diffeo},  in the $N$ levels case, $N>2$, the ordinary differential equation  \eqn{maineq}  will be replaced by partial differential equations. We plan to investigate the situation in higher dimensions elsewhere, by using the approach in terms of Hamilton-Jacobi theory \cite{CDFMMP}.

\begin{figure}[h!]
\centering
\includegraphics[scale=0.8]{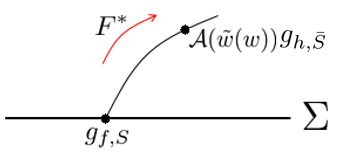}
\caption{\textit{The diffeomorphism (\ref{diffeo}) maps an element of the surface $\Sigma$ of Petz quantum metrics, associated to a generating function $S$ and an operator monotone function $f$, to a different quantum metric of the Petz type, characterized by another generating function $\bar S$ and by a positive and real analytic function $h$, times a conformal factor $\mathcal A\bigl(\tilde w (w)\bigr)$ which depends only on the change of the tomographic scheme.}}
\label{diff}
\end{figure}

Concerning the conformal factor, we argued that it naturally arises in the tomographic setting since a change of the tomographic scheme uniquely identifies a diffeomorphism on the parameter space whose action on the metric tensor defined on tomographic probabilities maps Fisher metrics into conformal Fisher metrics. Let us stress once again that this does not contradict Chenstov's uniqueness theorem since the diffeomorphism involves only the parameters of the tomographic probabilities and not the sample space. Similarly, on the  space of states  quantum metrics manifest an  analogue behavior under the action of such a diffeomorphism as sketched in fig. \ref{diff}.

Indeed, we have seen that the diffeomorphism identified by the change of tomography maps quantum Petz metrics into a different quantum Petz metric multiplied by the same conformal factor. Therefore, since the conformal factor depends only on the specific diffeomorphism and the latter is uniquely determined by the selected tomography, we conclude that the choice of a tomographic scheme selects a  function and hence a unique element in the family of Petz metrics according to the following diagram:

\begin{equation}
\xymatrix{
\widetilde G=\ar@{->}[dd]_-{\tilde w=\pm\sqrt{1-\widetilde G_{\tilde w\tilde w}^{-1}}\,}G_{\widetilde{\mathcal W}}^{(\text{Fisher})}\,\ar@{-->}[rr]^-{F^*}& & \,F^*\widetilde G=\mathcal A_F\,G_\mathcal W^{(\text{Fisher})}\,\ar@{->}[dd]^-{\,w=\pm\sqrt{1- G_{ww}^{-1}}}\\
& & \\
\tilde g=\tilde g_{S,f}^{(\text{Petz})}\,\ar@{-->}[rr]_-{F^*}& & \,F^*\tilde g=\mathcal A_F\,g_{\bar S,h}^{(\text{Petz})}\\
}
\end{equation}


\titleformat{\section}{\normalfont\Large\bfseries}{\appendixname~\thesection :}{0.5em}{}
\begin{appendices}
\section{Explicit computation of relation (\ref{erho})}		
\label{A}
We want to obtain the relation (\ref{erho}), $\tilde w=\tilde w(w)$, in the case of a tomography defined as
\begin{equation}
W_F(u|m_w)=\braket{m_w|uF(\rho)u^\dagger|m_w}
\end{equation}
where 
\begin{equation}
F(\rho)=\frac{e^{-\beta\rho}}{Tr(e^{-\beta\rho})}\;.
\end{equation}
Since both $F(\rho)$ and $\rho$ are states, they admit a Bloch decomposition
\begin{equation}
\rho=\frac{1}{2}(\sigma_0+w\sigma_w)\qquad,\qquad F(\rho)=\frac{1}{2}(\sigma_0+\tilde w\sigma_w).
\end{equation}
Then, we have
\begin{equation}
\label{fexp}
e^{-\beta\rho}=e^{\frac{-\beta\sigma_0}{2}}e^{\frac{-\beta w\sigma_w}{2}}=e^{-\frac{\beta}{2}}\begin{pmatrix}
\cosh\left(\frac{\beta w}{2}\right)-\sinh\left(\frac{\beta w}{2}\right)&0\\
0&\cosh\left(\frac{\beta w}{2}\right)+\sinh\left(\frac{\beta w}{2}\right)
\end{pmatrix}\;.
\end{equation}
The trace of this matrix is
\begin{equation}
Tr(e^{-\beta\rho})=2e^{-\frac{\beta}{2}}\cosh\left(\frac{\beta w}{2}\right).
\end{equation}
Therefore, by dividing Eq. (\ref{fexp}) for its trace, we obtain the matrix expression for $F(\rho)$
\begin{equation}
F(\rho)=\frac{1}{2}\begin{pmatrix}
1-\tanh\left(\frac{\beta w}{2}\right)&0\\
0&1+\tanh\left(\frac{\beta w}{2}\right)
\end{pmatrix}.
\end{equation}
By comparing it with its Bloch decomposition
\begin{equation}
F(\rho)=\frac{1}{2}\begin{pmatrix}
1+\tilde w&0\\
0&1-\tilde w
\end{pmatrix}\;,
\end{equation}
it follows that
\begin{equation}
1-\tanh\left(\frac{\beta w}{2}\right)=1+\tilde w\quad\Longrightarrow\quad\tilde w(w)=-\tanh\left(\frac{\beta w}{2}\right)\;,
\end{equation}
which is actually Eq. (\ref{erho}).

\bigskip

\noindent{\bf Acknowledgements} 

\noindent We are  indebted with F. M. Ciaglia and F. Di Cosmo for discussions and remarks during the revision  of the manuscript. G.M.  would like to acknowledge the grant ``Santander-UC3M Excellence Chairs 2016. P.V.  acknowledges  support by COST (European Cooperation in Science  and  Technology)  in  the  framework  of  COST  Action  MP1405  QSPACE.
\end{appendices}

  \end{document}